\documentclass[aps,twocolumn,10pt]{revtex4}
\usepackage{graphicx,amssymb,amsmath,subfigure}

\newcommand{\braket}[1]{\left<#1\right>}
\newcommand{\para}[1]{\left(#1\right)}
\newcommand{\abs}[1]{\left|#1\right|}

\begin{document}

\title{Self-consistent calculation of the single particle scattering rate in high $Tc$ cuprates}
\author{Abolhassan Vaezi}
\affiliation{Department of Physics,
Massachusetts Institute of Technology,
Cambridge, MA 02139, USA}
\email{vaezi@mit.edu}

\date{\today}

\begin{abstract}
The linear temperature dependence of the resistivity above the optimal doping is a longstanding problem in the field of high temperature superconductivity in cuprates. In this paper, we investigate the effect of gauge fluctuations on the momentum relaxation time and the transport scattering rate within the slave boson method. We use a more general slave treatment to resolve the ambiguity of decomposing the Heisenberg exchange term. We conclude that this term should be decomposed only in the Cooper channel. This results in the spinon mass inversely proportional to the doping. It is showed that solving the equation for the transport scattering rate self-consistently, we find a crossover temperature above which we obtain the linear temperature dependence of the electrical resistivity as well as the single particle scattering rate. It is also shown that this linear temperature dependence of the scattering rate in the pseudogap region explains the existence of the Fermi arcs with a length that linearly varies with temperature.
\end{abstract}

\maketitle

\section{Introduction}
It has been emphasized by many authors that the physics of high temperature superconductivity in cuprates \cite{Bednorz_Mueller_1986} should be viewed as a doped Mott insulator \cite{Anderson_1987Sci,Lee_Nagaosa_Wen_2006a,Senthil_2009a}. This proximity to the Mott insulating phase at half filling, sheds some light on the underlying pairing mechanism of the superconductivity. One of the useful and rather successful treatments of the doped Mott insulator is the slave particle method \cite{Coleman_1983_1,Anderson_Zou_a,Vaezi_2010b}. It is well-known that this method results in the emergent internal gauge field that is strongly coupled to slave particles. Indeed it is the dressed slave particles by gauge bosons that is physical and observable.

Since slave particles are strongly interacting with gauge bosons, the effect of gauge interactions on the physical properties of electrons should be taken into consideration in any realistic model. This effect has been extensively studied in the past three decades. For example, Ioffe and Larkin \cite{Ioffe_1} have studied how the physical properties of electrons, {\em e.g.} their electrical conductivity, are related to the corresponding properties of slave particle, in the presence of the gauge field. Also, many authors \cite{PA_Lee_1992_1,Senthil_2009a} have studied the effect of gauge fluctuations on the transport properties of electrons. Here we follow their method but with two modifications. First of all, we argue that the mass of spinons is inversely proportional to the doping, instead of being almost independent of it. As a result, the Fermi velocity of electrons should scale with doping outside the superconducting phase. The second difference is that we solve equations self-consistently. We show that either a simple scaling argument or exact numerical calculation, results in the transport scattering time that is linear in T above a crossover temperature, $T^*$ that scales with the doping.

\section{ Linear temperature dependence of the resistivity} From the Fermi liquid theory we expect $T^2$ dependence for the electrical resistivity. This reflects the stability of this phase at low temperatures. On the other hand, in the normal state above the optimal doping (strange metal phase of the hole doped cuprates), the resistivity exhibits a linear dependence on temperature instead of the expected parabolic dependence \cite{Bednorz_Mueller_1986,Lee_Nagaosa_Wen_2006a,Yan_2008_1}. For temperatures comparable to the Debye temperature and higher, this linear dependence can be explained through electron phonon interaction. The Debye temperature is several hundred Kelvin while this behavior remains down to much smaller temperatures. So this is a fundamental issue that cannot be explained only through simple electron phonon interaction. In slave boson approach, holons and spinons are strongly interacting with the internal gauge field. One popular notion is that this interaction can explain the desired behavior. By calculating the self energy of electrons due to scattering off the gauge field at finite temperature, we can compute the transport lifetime, as well as the momentum and energy relaxation time of quasiparticles resistivity.

\begin{figure}
\centering
\subfigure{
\includegraphics[width=210pt]{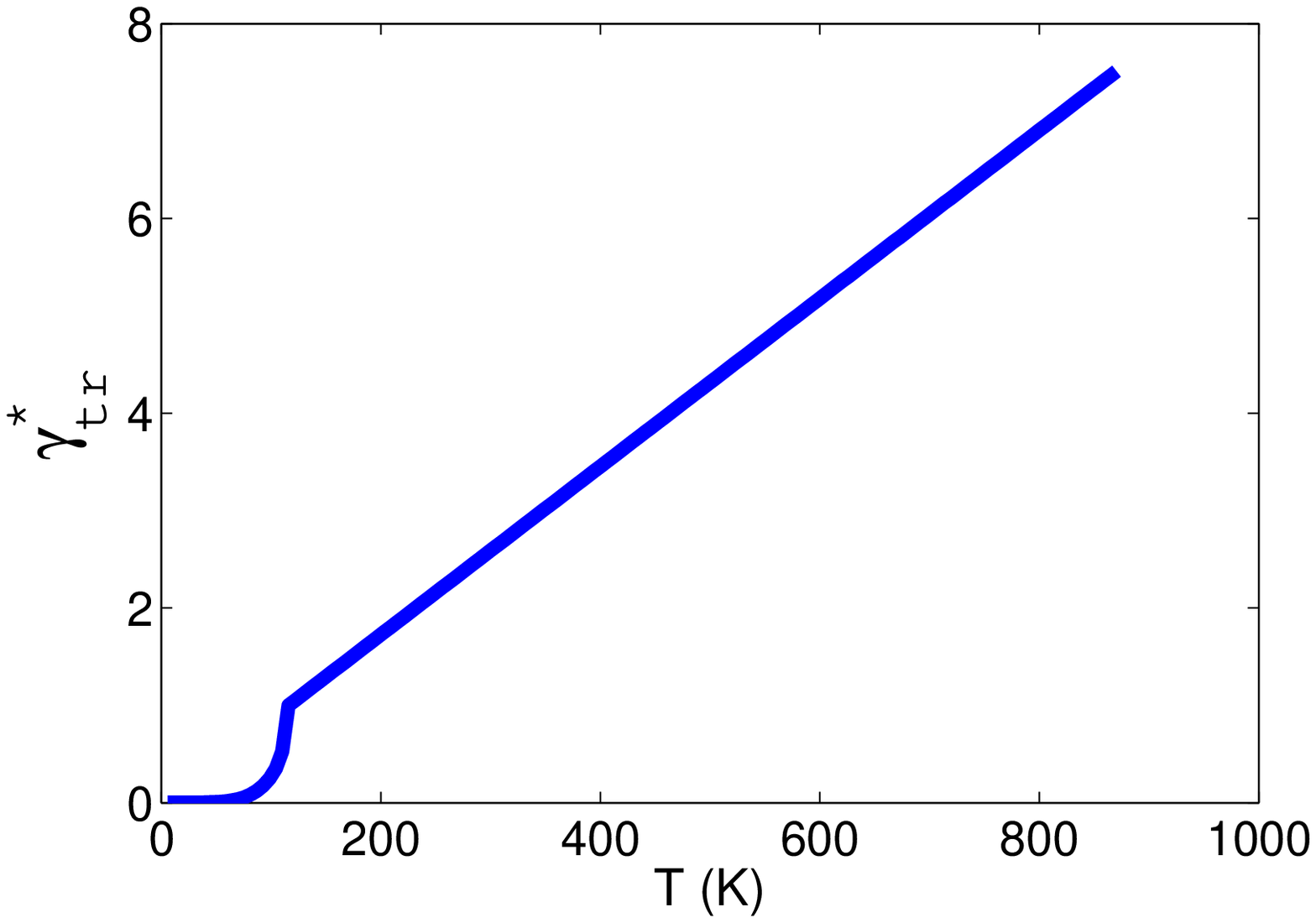}
\label{fig:subfig1a}
}
\subfigure{
\includegraphics[width=210pt]{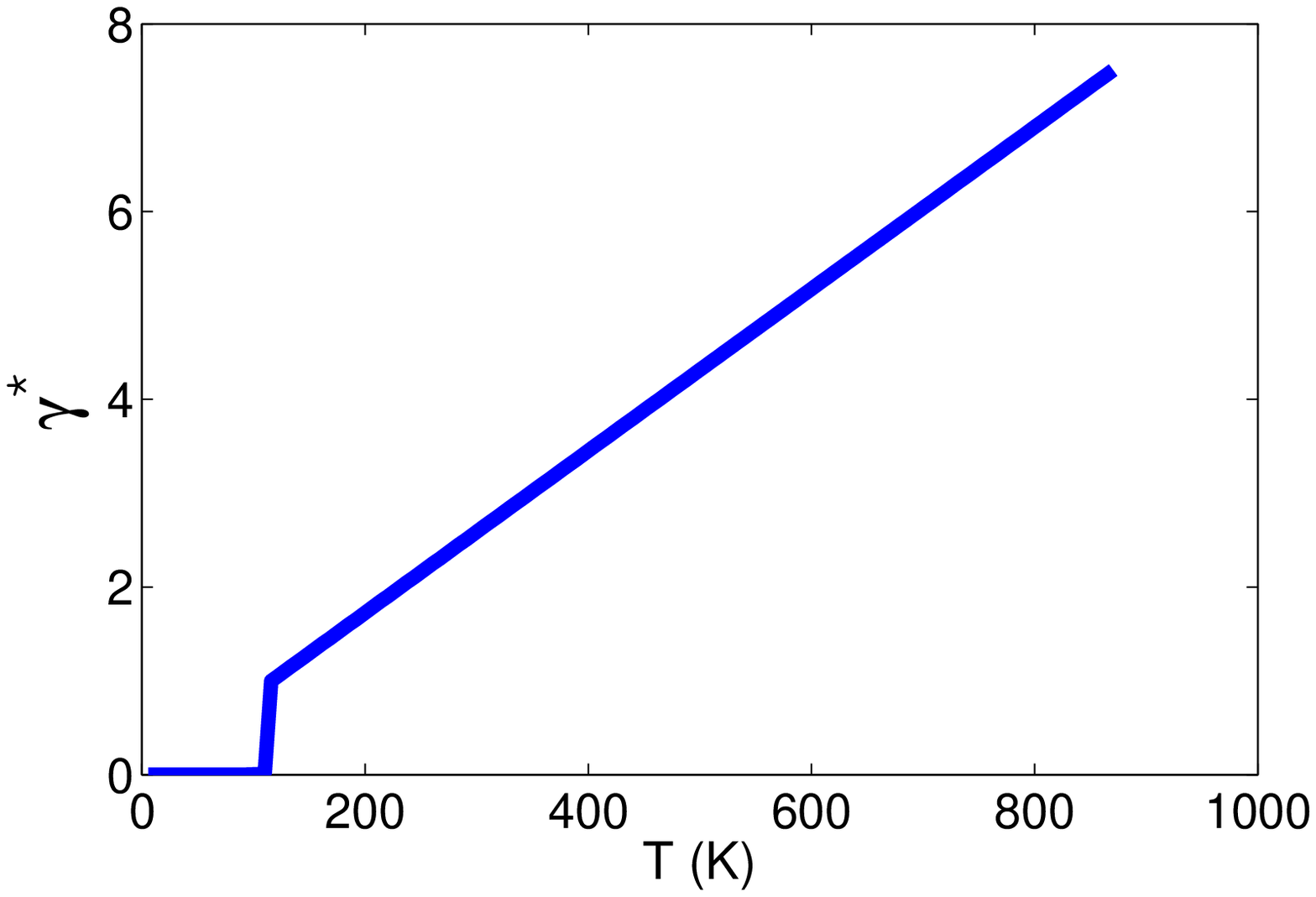}
\label{fig:subfig1b}
}
\label{fig:subfigureExample}
\caption[Optional caption for list of figures]{{\bf Scattering rate.---} Self-consistent calculation of the transport scattering rate $\gamma_{tr}$ and momentum relaxation scattering rate $\gamma$ versus temperature in the absence of external magnetic field. We assume that Cooper pairs are killed so we can study the crossover between the Fermi liquid and the strange metal phases. At low enough temperatures, holon gas condenses and we obtain FL phase. At high enough temperatures, we obtain linear temperature dependence for $\gamma_{tr}$ and $\gamma$ signaling the strange metal phase. \subref{fig:subfig1a}, Reduced inverse transport lifetime $\gamma^{*}_{tr}=\frac{\gamma_{tr}\para{T}}{\gamma_{tr}\para{T^*}}$ at $x=0.15$. Temperature is in units of Kelvin. Below the crossover temperature $T^*\sim 100 K$, holons undergo Bose Einstein condensation and we obtain Fermi liquid behavior, $\gamma_{tr}\sim T^2$. Above  $T^*$, we obtain marginal non-Fermi liquid with $\gamma_{tr}\sim T$. Interestingly $T^*$ scales with doping $x$, like the BEC transition temperature of the holon gas. \subref{fig:subfig1b},
 Reduced inverse momentum relaxation time $\gamma^{*}=\frac{\gamma\para{T}}{\gamma\para{T^*}}$ at $x=0.15$.}
\end{figure}\label{Fig1}

Some authors have studied the effect of gauge fluctuations on the transport properties of cuprates, and they find $T^{4/3}$ temperature dependence for the resistivity \cite{PA_Lee_1992_1,Senthil_2009a}. Here we closely follow their method, but with two important differences. First of all they assume the mass of spinons has a weak dependence on the doping and is inversely proportional to the exchange energy $J$. Using the more general Anderson-Zou slave boson method \cite{Anderson_Zou_a,Vaezi_2010b}, we obtain a different behavior. In this method, the spinon mass is inversely proportional to the doping. This for instance means that the Fermi velocity outside the superconducting phase {\em e.g.} in the Fermi liquid phase, is proportional to the doping. The doping dependence of the spinon mass should be observed in the density of states at Fermi level. The second difference is that the problem should be solved self-consistently. In the rest of this paper, we have provided the details of calculation and we have presented an integral expression for the transport scattering rate $\gamma_{tr}$ that depends on itself. If we keep $\gamma_{tr}$ inside the integral, along with the assumption $1/m^*_s\sim x$, we obtain a temperature scale $T^*$ above which we find linear temperature resistivity. Surprisingly $T^*$ scales with doping $x$, the same behavior that we expect for the crossover temperature between the Fermi liquid and strange metal phases. Below this temperature scale, holons condense and we recover the results of the Fermi liquid theory \cite{Sachdev_2004_1}. We have solved the self-consistent equation for $\gamma_{tr}$ and we verify $T^2$ behavior for the resistivity at low temperatures in the Fermi liquid phase and linear temperature resistivity above $T^*$ in the strange metal phase. Fig. 1 summarizes our result.

\section{Method}

Hubbard model is the simplest model that captures the physics of the Mott insulators. The Hamiltonian of the Hubbard model is defined as,

The Hubbard model is defined as the following:
\begin{eqnarray}
H=U\sum_in_{i,\uparrow}n_{i,\downarrow}-t\sum_{\left<i,j\right>,\sigma}c_{j\sigma}^\dag
c_{i\sigma}+h.c.  \end{eqnarray}

Here $\braket{i,j}$ means site $j$ is one of the nearest
neighbors of site $i$.


Now let us employ a more general slave boson approach to the Hubbard model \cite{Senthil_2009a,Anderson_Zou_a,Vaezi_2010b}. In this approach an electron operator decomposes in the following way

\begin{eqnarray}
  c_{i,\sigma}=f_{i,\sigma}h_{i}^\dag+\sigma f_{i,-\sigma}^\dag d_{i}\label{Eq35}
\end{eqnarray}

where $f_{i,\sigma}^\dag$ is spinon creation operator (assumed to be fermion), $h_{i}^\dag$ is holon creation operator and $d_{i}^\dag$ is doublon creation operator. We assume spinons are fermions and obey Fermi Dirac statistics and are electrically neutral, while holons and doublons are bosons and have $+e$ and $-e$ electric charges respectively. Spin $\sigma$ spinon corresponds to a site with only one electron with spin up, holon represents an empty site and doublon corresponds to a doubly occupied site. As it is clear from the above definition, electron operator is invariant an under internal compact U(1) gauge transformations ({\em i.e.} if we multiply each creation operator by a local phase $exp(i\theta_{i})$), provided all slave particles carry the same internal gauge charge({\em i.e.} $\theta_{i}$ is the same for all slave particles at site $i$). Note that we must keep the bosonic statistics of holons and doublons after gauge transformation, otherwise we had SU(2) gauge freedom.  In order to have fermionic statistics for electron operator, we have to implement $f_{i,\uparrow}^\dag f_{i,\uparrow}+f_{i,\downarrow}^\dag f_{i,\downarrow}+h_{i}^\dag h_{i}+d_{i}^\dag d_{i}=1$ constraint at each site. On the other hand, doping implies $\braket{h_{i}^\dag h_{i}-d_{i}^\dag d_{i}}=x$ constraint.

Since in the Mott insulator where onsite Coulomb repulsion is very large ($U\ll t$), the double occupancy is very costly, and these states can be removed in the low energy studies of cuprates. In other words, for large $U/t$ limit of the Hubbard model, charge excitation gap is of order $U$ and therefore we can integrate out doublons to obtain an effective action for spinon and holons. This process provides a systematic way to recover the t-J model and naturally reduces to the famous slave boson theory of the t-J model.

In the t-J model, there is at most one electron at each site. To implement this constraint in the slave boson method, at we discussed before, empty states are treated as the charged bosonic particles, dubbed as holon. So we can take the non-doubly occupancy constraint by using the Lagrange multiplier method. The physical Hilbert space, contains three states at each site, occupied state with spin up or down and unoccupied state (empty sites). However, the Hilbert space of the slave boson method is much larger as we can have as many holons per site as we want and the constraint is implemented only in average. Therefore the meanfield description of the slave boson method is incomplete and redundant. In the next section, it is shown that this redundancy is responsible for the emergent $U(1)$ gauge field. Although at the beginning there is no kinetic term for the gauge field, upon renormalization and by integrating out the slave particles, this term will be generated and the gauge field will have its own dynamics in that case. Slave particles are interacting strongly with this gauge field and their scattering off gauge potential gives them a finite lifetime. The single particle scattering rate is computed in section VI and we show that it is results in a transport scattering rate that is linear in temperature.

One ambiguity in the t-J model is how to decompose the exchange term $JS_{i}.S_{j}$, where $S_{i}$ is the spin operator. Most authors, decompose this term in direct and Cooper channels symmetrically. Within meanfield approximation, they rewrite  $J\sum_{i,j}S_{i}.S_{j}$ as

\begin{eqnarray}
  && -3/8J \sum_{i,j,\sigma}\braket{\chi_{i,j}^{f}} f_{i,\sigma}^\dag f_{j,\sigma}+H.c.\cr
  && -3/8J \sum_{i,j,\sigma}\braket{\Delta_{i,j}^{f}} \sigma f_{i,\sigma}^\dag f_{j,-\sigma}^\dag +H.c.\cr
  && +3/8J \sum_{i,j}\para{\abs{\braket{\chi_{i,j}^{f}}}^2+\abs{\braket{\Delta_{i,j}^{f}}}^2}
\end{eqnarray}
where $\chi_{i,j}^{f}=\sum_{\sigma}f_{i,\sigma}^\dag
f_{j,\sigma}$ and $\Delta_{i,j}^{f}=\sum_{\sigma}\sigma f_{-\sigma,i}
f_{j,\sigma}$. At small dopings, the above hopping implies $1/m^*\sim J \chi_{i,j}$ which has a weak dependence on the doping. However, if we use the more general slave boson approach, it resolves the above mentioned ambiguity and we have only one choice. In this formalism, the exchange term only decomposes in the Cooper channel. Using the Eq. [2], within the meanfield approximation, the Hubbard model can be rewritten as

\begin{eqnarray}
&&H=\sum U d_{i}^\dag
d_{i}-t\sum_{\braket{i,j}}\para{\chi_{i,j}^{s}\chi_{j,i}^{b}+\Delta_{i,j}^{s}\Delta_{i,j}^{b }+H.c.}~~~~
\end{eqnarray}
in which we
have used these notations
$\chi_{i,j}^{b}=h_{i}^\dag h_{j}-d_{i}^\dag
d_{j}$ and $\Delta_{i,j}^{b}= d_{i}h_{j}+h_{i}d_{j}$. By integrating out doublons, it can be shown that $\braket{\chi_{b}}\sim x$, and $\braket{\Delta_{b}}\sim t/U$. $\Delta_{i,j}^{s}\Delta_{i,j}^{b }$ represents the exchange term and after integrating out doublons it decouples only in the Cooper channel (we replace the exchange term by $\braket{\Delta_{i,j}^{b }}\Delta_{i,j}^{s}\sim \frac{t}{U}f^\dag f^\dag+H.c.$ form). Therefore the hopping term for spinons is $-t \braket{\chi_{b}} \sum_{i,j} f_{i,\sigma}^\dag f_{j,\sigma}$ and as results they have a mass $1/m_{s}^{*}\sim 2tx$. As we approach the half filling (undoped material), the effective mass of spinons diverges signalling the metal-Mott insulator phase transition.

So we end up at the the t-J model starting from the large U limit of the Hubbard model. It is believed that the t-J model captures the essential physics of the strongly correlated systems. This model is defined as the following

\begin{eqnarray}
  H_{t-J}=-t\sum_{\left<i,j\right>, \sigma }P_{G}c_{\sigma,i}^\dag c_{\sigma,j}P_{G}+J\sum_{i,j}\hat{S}_{i}.\hat{S}_{j}
\end{eqnarray}

where $P_{G}$ is the Gutzwiller projection operator that removes doubly occupied states. Within the slave boson formalism, after removing doubly occupied states, electrons can be decomposed as $c_{i,\sigma}^\dag = f_{i,\sigma}^\dag h_{i}$ along with the physical constraint on each site:  $h_{i}^\dag h_{i}+\sum_{\sigma}f_{i,\sigma}^\dag f_{i,\sigma}=1$ which implements the Gutzwiller projection. Whenever $c_{i,\sigma}^\dag$ acts on an empty site, it annihilates one holon and creates a spinon with spin $\sigma$. We cannot act further on the resulting state by $c_{i,-\sigma}^\dag$, since this operator has to kill a holon, but there is no holon anymore at that site. If we act $c_{i,\sigma}$ on a site that contains a spinon with spin $\sigma$, the operators annihilates the spinon and creates a holon at that site. So by acting projected electron operator we always annihilate one type of slave particle and create another one and therefore the number of slave particles at each site is conserved. Now we can rewrite the t-J model in terms of the new slave particles. Within meanfield approximation and by using Hubbard-Stratonovic transformation, we can decouple spinons (spin sector) from holons (charge sector) and we obtain the following effective Hamiltonians for each sector

\begin{eqnarray}
  &&H_{h}= -\sum_{<i,j>}t\chi_{s} h_{i}^\dag h_{j}-\sum_{i}\mu_{h}h_{i}^\dag h_{i}\\
  &&H_{s}=-\sum_{<i,j>,\sigma}t\chi_{h}f_{i,\sigma}^\dag f_{j,\sigma} -\sum_{i,\sigma}\mu_{s}f_{i,\sigma}^\dag f_{i,\sigma}\cr &&-\sum_{<i,j>}~\para{J/2}\Delta_{s}\para{i,j}\para{f_{i,\uparrow}^\dag f_{j,\downarrow}^\dag-f_{i,\downarrow}^\dag f_{j,\uparrow}^\dag}  +H.c. ,~~
\end{eqnarray}
where the following notations have been used

\begin{eqnarray}
&&  \chi_h=\braket{h_{i+\vec{\delta}}^\dag h_{i}}\\
&&  \chi_s=\braket{\sum_{\sigma}f_{i+\vec{\delta},\sigma}^\dag f_{i,\sigma}}\\
&&  \Delta_{s}\para{i,j}=\frac{1}{2}\braket{f_{i,\uparrow}^\dag f_{j,\downarrow}^\dag-f_{i,\downarrow}^\dag f_{j,\uparrow}^\dag}
\end{eqnarray}

At low temperatures, most of holons occupy the groundstate with momentum $k=0$, therefore $\chi_{h}\sim x$. This model has been extensively studied in the literature and it is well known that this model leads to the d-wave pairing symmetry for spinons \cite{Lee_Nagaosa_Wen_2006a}, {\em i.e.} $\Delta_{s}\para{\pm \hat{x}}=\Delta_{s}$ and $\Delta_{s}\para{\pm \hat{y}}=-\Delta_{s}$.

\section{Single particle scattering rate}

Now, we discuss the effect of gauge fluctuations on the transport properties of cuprates in the strange metal phase. Here we closely follow the method used by Lee and Nagaosa \cite{PA_Lee_1992_1} and a more recent approach by Senthil and Lee \cite{Senthil_2009a}.


In the continuum approximation, the hopping part of the action of spinons and holons are written as

\begin{eqnarray}
&&~~~ \int dx dt f_{\sigma}^\dag\para{x} \para{-i\partial_{t}-\nabla ^2/2m^*_s -\mu_{s}} f_{\sigma}\para{x} \cr
&& + \int dx dt h^\dag \para{x} \para{-i\partial_{t}-\nabla ^2/2m^*_h -\mu_{s}} h\para{x}
\end{eqnarray}

As we pointed out, slave boson formalism has U(1) gauge theory. Under the gauge transformation, $f_{i,\sigma}\rightarrow e^{i\theta_i}f_{i,\sigma}$ and $h_{i}\rightarrow e^{i\theta_i}h_{i}$. So $\chi^{s,h}_{i,j}\rightarrow e^{i\para{\theta_j-\theta_i}}\chi_{i,j}^{s,h}$. If we define $a_{\vec{\delta}}\para{i}$ as $\chi_{i,j}=\abs{\chi_{i,j}}e^{ia_{\vec{\delta}}\para{i}}$, where $\vec{\delta}=\vec{R}_{i}-\vec{R}_{j}$, we have $a_{\vec{\delta}}\para{i}\rightarrow a_{\vec{\delta}}\para{i}+\theta_{j}-\theta_{i}$. This implies that in the continuum model, $a_{\mu}\para{x} \rightarrow a_{\mu}\para{x}-\partial_{\mu} \theta\para{x}$. Therefore we should add modify the continuum model in the following way to obtain a gauge invariant model

\begin{eqnarray}
&&~~~ \int dx dt f_{\sigma}^\dag\para{x} \para{-iD_{t}-D_{i}^{2} /2m^*_s -\mu_{s}} f_{\sigma}\para{x} \cr
&& + \int dx dt h^\dag \para{x} \para{-iD_{t}-D_{i}^{2}/2m^*_h -\mu_{s}} h\para{x} \end{eqnarray}
where $D_{\mu}=\partial_{\mu}-ie_{int} a_{\mu}/c$, where $e_{int}$ is the internal gauge charge. We scale $a_{\mu}$ so that $e_{int}=e$.

To obtain an effective action for gauge particles we can integrate out spinons and holons. By expanding $D_{\mu}$ in terms of $\partial_{\mu}$ and $a_{\mu}$, it is clear that in the continuum model, the vector potential is minimally coupled to matter field, {i.e.} the gauge field is coupled to the current carried by quasiparticles and we have the following interaction between gauge field and spinons
\begin{eqnarray}
  e\int dxdt J_{s}^{\mu}\para{x,t}a_{\mu}\para{x,t} \end{eqnarray}
where $a^{\mu}=\para{\phi, a_x,a_y}$ and $J^{\mu}=\para{n, J_{x},J_{y}}$, where $n_{s}=f^\dag f$, $\vec{J}_{s}=i f^\dag \para{\vec{\nabla}/2m_{s}} f+H.c.$ . We have a similar term for holons. Now we use the gauge freedom to set $\phi=0$ and choose the Coulomb gauge $\vec{\nabla}. \vec{a}=0$. Up to the second order perturbation theory the above interaction terms leads to an following effective action for the gauge field proportional to the following form,
\begin{eqnarray}
  \int dx d\tau dx'd\tau' a_{i}\para{x,\tau}\braket{J^{i}\para{x,\tau},J^{j}\para{x',\tau'}}a_{j}\para{x,t} \end{eqnarray}

This is only the contribution coming from paramagnetic response. There is a similar contribution from the diamagnetic part. Lee and Nagaosa have studied this problem in detail. Following their method, we can find the gauge field propagator $D_{\mu,\nu}\para{r,\tau}=\braket{T_{\tau}a_{\mu}\para{r,\tau}a_{\mu}\para{r',0}}$. it can be shown that the gauge field propagator can be written as $D_{ij}\para{q}=\para{\Pi_{s}+\Pi_{h}}_{ij}^{-1}$, where
\begin{eqnarray}
  \Pi_{s}^{ij}=-\braket{T_{\tau}\left[J_{s}^{i}\para{r,\tau}J_{s}^{j}\para{0,0}-\delta_{ij}n_{s}\delta\para{r}\delta\para{\tau}\right]} \end{eqnarray}
and similarly for $\Pi_{h}^{ij}$. In the Coulomb gauge the spatial part of the gauge field is transverse and can be written as
\begin{eqnarray}
D_{ij}\para{q,\omega}=\para{\delta_{ij}-q_i q_j /q^2} D^{T}\para{q,\omega} \end{eqnarray}
where we have defined $D^{T}\para{q,\omega}$ as
\begin{eqnarray}
D^{T}\para{q,\omega}=\left[\Pi^{T}_{s}\para{q,\omega}+\Pi^{T}_{h}\para{q,\omega}\right]^{-1} \end{eqnarray}
For small $q$ and $\omega$ we can use the following approximation
\begin{eqnarray}
\Pi_{s}^{T}\para{q,\omega}=i\omega\sigma_{s}^{T}\para{q,\omega}-\chi^{D}_{s}q^2-\rho_{s,c}\para{T} \end{eqnarray}
where $\sigma_{s}^{T}$ is the transverse conductivity and $\chi^{D}_{s}$ is the Landau diamagnetic susceptibility of the spinon system, which equals $1/24\pi m_s$ in 2D fermion systems. $\rho_{s,c}$ is the condensation fraction of the spinon system in the paired state and in the normal state it is zero. Similarly for holons we have
\begin{eqnarray}
\Pi_{s}^{T}\para{q,\omega}=i\abs{\omega}\sigma_{h}^{T}\para{q,\omega}-\chi^{D}_{h}q^2-m_{h}\rho_{h,c}\para{T} \end{eqnarray}
where $\chi^{D}_{h}=n\para{0}/48\pi m_{s}$ in which $n\para{0}$ is the occupation number of the ground state and $\rho_{h,c}$ is the superfluid density of holons, which is zero in the strange metal phase as well as the pseudogap phase. We see that the propagator of the gauge field that determines the transport scattering rate of spinons, depends on the optical conductivity of spinons. On the other hand, optical conductivity itself depends on the transport scattering rate, since we have $\sigma_{s}\para{q,\omega}=v_{_F} k_F/\sqrt{4\gamma_{tr}^2+v_{_F}^2 q^2}$. If we neglect $\gamma_{tr}$ in the $\sigma_{s}\para{q,\omega}$, we recover the results of the Ref. \cite{PA_Lee_1992_1}, except that since the spinon mass is much larger in our case, the crossover temperature is smaller. In Ref. \cite{PA_Lee_1992_1}, Lee and Nagaosa have calculated the scattering rate of holons due to interaction with the gauge field and they report a linear temperature behavior above the BEC transition. They have also studied the transport scattering rate of spinons and they report $T^{4/3}$ in the strange metal phase, which is slightly different from the linear temperature behavior.
Now let us calculate the self energy of spinons due to their interaction with the gauge field. Then by looking at its imaginary part we can deduce the scattering rate. According to Senthil and Lee, up to the first order approximation, the self energy of spinons due to their scattering off the gauge field is described by the following expression

\begin{eqnarray}
  &&\Sigma\mbox{\H{}}\para{K,\Omega=0}=-\pi\int d\omega dq \para{v_{_F}\times q}^2 D\mbox{\H{}}\para{q,\omega}\cr
  &&\times A\para{K-q,-\omega}\para{n_{_{BE}}\para{\omega}+n_{_{FD}}\para{\omega}} \end{eqnarray}
where $n_{_{BE}}$ and $n_{_{FD}}$ are Bose Einstein and Fermi-Dirac distribution functions respectively. Near the Fermi surface the above integral reduces to

\begin{eqnarray}
  \gamma=\pi v_{_F}K_{_F}\int d\omega \frac{1}{\sinh \beta \omega}\int dq D\mbox{\H{}}\para{q,\omega} \end{eqnarray}
where $D\mbox{\H{}}\para{q,\omega}=Im\para{i\omega \sigma^{T}_{q,\omega}-\chi^{D} q^2 -\rho_{c}}^{-1}$ in which $\sigma=\sigma_{s}+\sigma_{h}$, $\chi^{D}=\chi^{D}_{s}+\chi^{D}_{h}$ and $\rho_{c}=\rho_{c,s}+\rho_{c,h}$. To compute $\gamma_{tr}$ we should multiply $D\mbox{\H{}}\para{q,\omega}$ by $\para{\frac{q}{K_{_F}}}^2$. So we have

\begin{eqnarray}
  \gamma_{tr}=\pi v_{_F}K_{_F}\int d\omega \frac{1}{\sinh \beta \omega}\int dq \para{\frac{q}{K_{_F}}}^2 D\mbox{\H{}}\para{q,\omega}
\end{eqnarray}

Let us assume that $v_{_{F}}\sim K_{_F}/m^*_s \sim 2txK_{_F}$ and at small dopings, $K_{_F}\sim \pi$. When $q \ll \gamma_{tr}/v_{_F}$, $\sigma_{s}\para{q,\omega}=v_{_F} k_F/\gamma_{tr}$. On the other hand, in the absence of holon and spinon condensation, $\omega \sigma \para{q,\omega} \sim \chi^{D}q^2$, so $q \sim K_{F}\para{\frac{\omega}{2\gamma_{tr}}}^{1/2}$. Therefore the $q < q^{*}\sim \gamma_{tr}/v_{_F}$ region contribution to the integral over $q$ is of order $K_{_F}\para{\gamma_{tr}/v_{_F}}^{1/2}$. For $q^{*}$ comparable to a fraction of $\pi$ this is the leading order contribution. This corresponds to $\gamma^{tr}\para{T}\propto v_{_F}\propto tx$. As long as this is satisfied we have

\begin{eqnarray}
  \gamma^{3/2}_{tr} \propto \pi v_{_F}K_{_F}^{2}\int d\omega \frac{\omega^{1/2}}{\sinh \beta \omega}
\end{eqnarray}

A simple scaling argument verifies that $\gamma_{tr}\sim T$ satisfies the above self-consistent equation for $\gamma_{tr}$. The crossover temperature can be found from $\gamma^{tr}\para{T^{*}}\sim v_{_F}\propto tx$. Therefore $T^{*}\propto tx$, which is the expected behavior for the crossover temperature between the Fermi liquid phase and the strange metal phase.

Numerically, we solved this self-consistent equation exactly and we again we obtained linear $T$ dependence of $\gamma_{tr}$ above a temperature scale $T^*\para{x}$ that scales with doping $x$ (see Fig. 1). On the other hand the BEC transition temperature of holons scales with doping as well. In summary, above a temperature scale comparable to the Bose Einstein transition temperature of holon gas, we obtain linear temperature transport scattering rate. It should be mentioned that the scattering rate of holons is linear in $T$ even in the previous calculations \cite{PA_Lee_1992_1}. To obtain the physical quantities from the corresponding quantities for spinons and holons, we should recombine them in a particular way, that is called Ioffe-Larkin formula. This formula tells us that the physical conductivity is related to that of holons and spinons in the following way $\sigma^{-1}=\sigma_{s}^{-1}+\sigma_{h}^{-1}$. For dc conductivity we have $\sigma=ne^2\tau_{tr}/m$ and $\tau_{tr}=1/\gamma_{tr}$. Since the scattering rate of holons and spinons are both linear in $T$, the dc conductivity is is linear in temperature as well. In Fig. 1 we have presented our numerical results.

On the other hand, below the BEC transition temperature, $\rho_{c,h}\neq 0$. Numerically, we obtain $T^2$ behavior for the $\gamma_{tr}$ in this phase, which is the right sign and is expected for the Fermi liquid phase.

\begin{figure}
\centering
\subfigure{
\includegraphics[height=165pt,width=190pt]{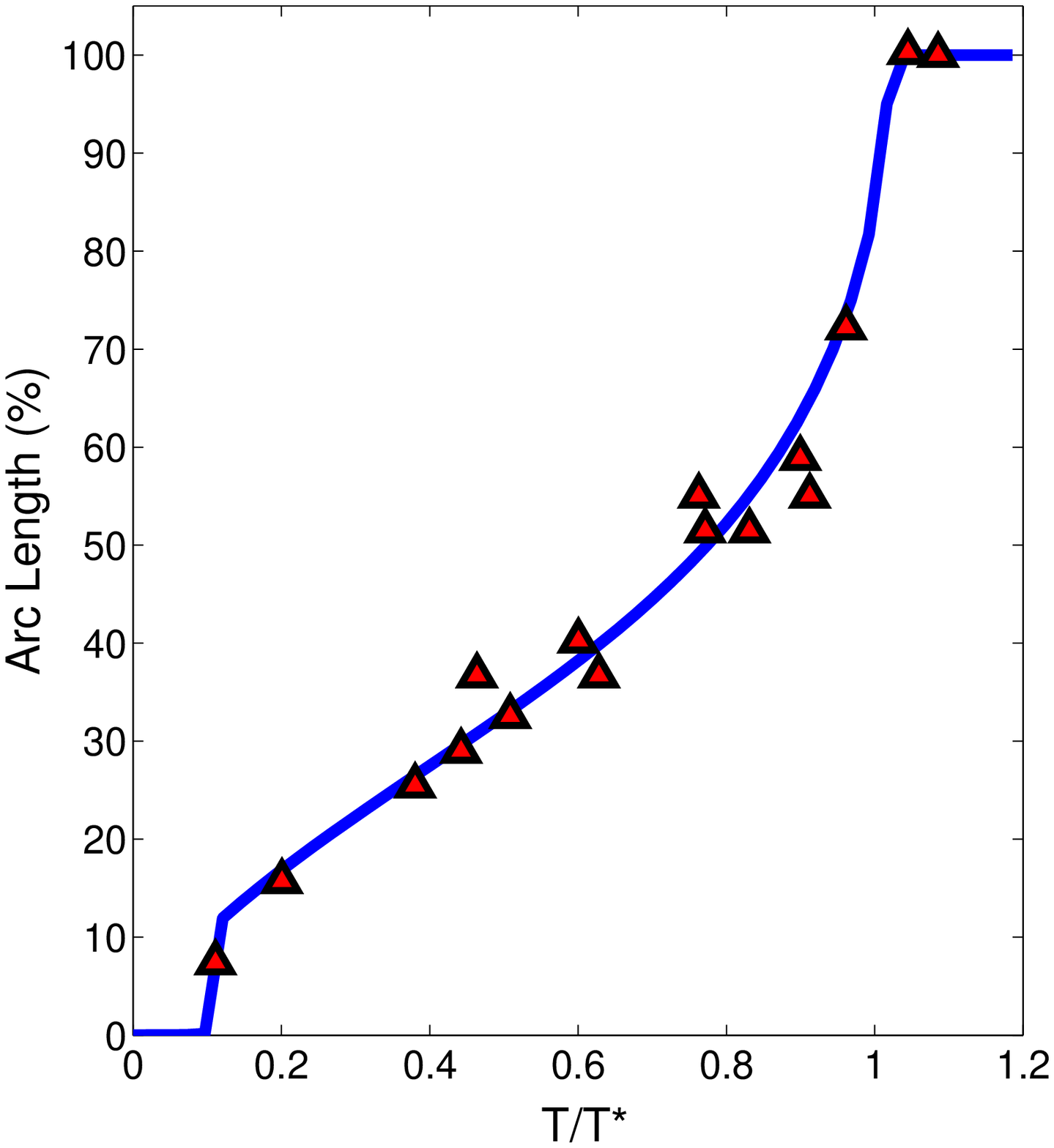}
\label{fig:subfig2a}
}
\subfigure{
\includegraphics[height=165pt,width=190pt]{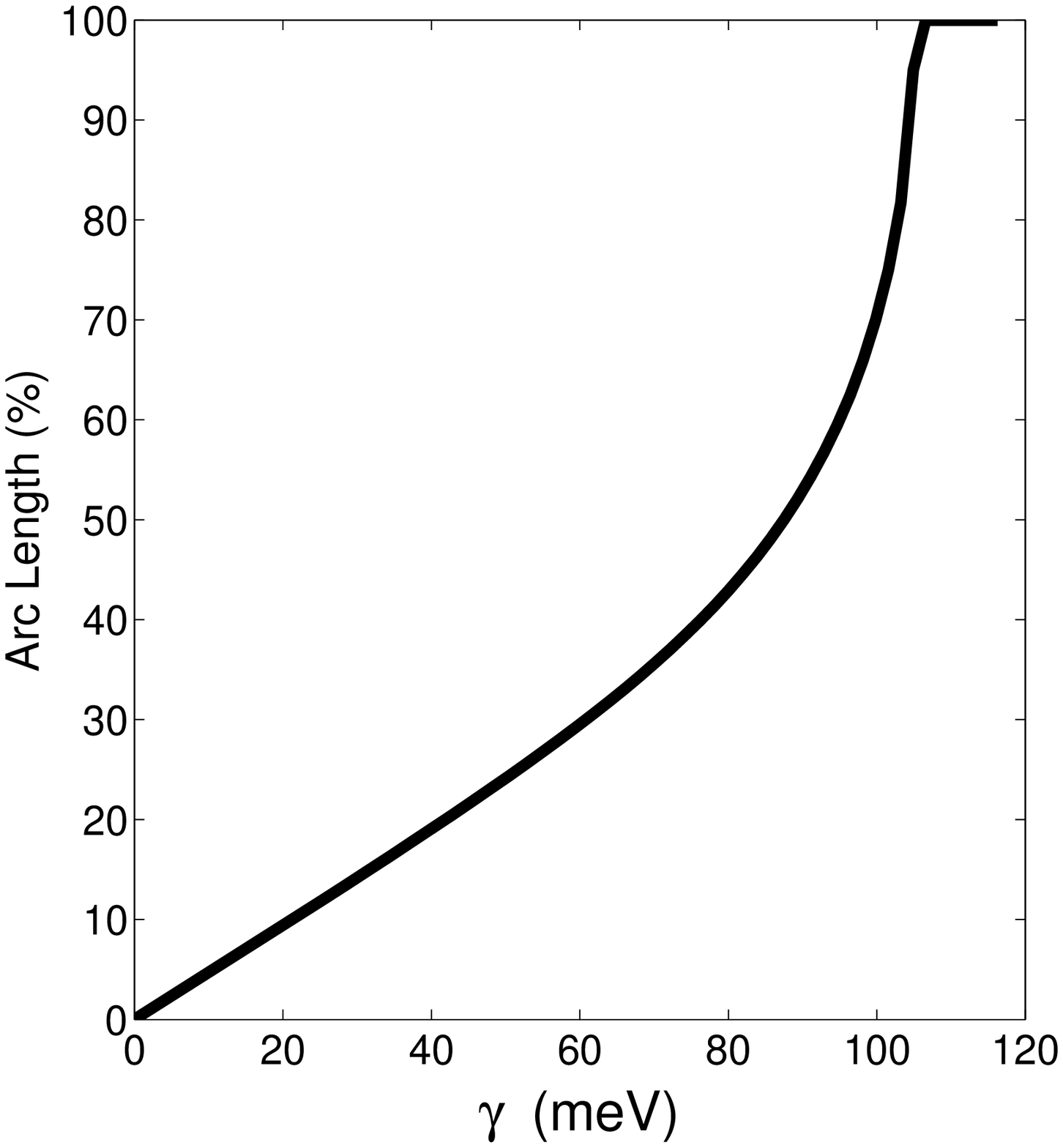}
\label{fig:subfig2b}
}
\label{fig:subfigureExample}
\caption[Optional caption for list of figures]{{\bf Length of the Fermi arc.---} Length of the Fermi arc. In the pseudogap region, we obtain disjoint Fermi segments, at which excitations are gapless. The length of these segments increases with temperature and the scattering rate $\gamma$. \subref{fig:subfig2a}, The length of the Fermi arc vs. normalized temperature ($T/T^*$). The blue line our theoretical calculation and red triangles are experimental data taken from Ref. \cite{Kanigel_2006_a}. Below the superconducting transition temperature $T_c$, the scattering rate $\gamma$ is very small due to the condensation of quasiparticles. Therefore only very near nodal points we have gapless excitation. Above $T_c$, there is macroscopic condensation and the scattering rate varies linearly with temperature. As $\gamma$ increases, the length of the Fermi arc increases. At $T^*$, we obtain closed Fermi surface. \subref{fig:subfig2b}, The length of the Fermi arc vs. scattering rate $\gamma$. Below the superconducting transition temperature $T_c$, the scattering rate $\gamma$ is very small due to the condensation of quasiparticles. Therefore only very near nodal points we have gapless excitation. Above $T_c$, there is macroscopic condensation and the scattering rate varies linearly with temperature. As $\gamma$ increases, the length of the Fermi arc increases. At $T^*$, we obtain closed Fermi surface. }\label{fig2}
\end{figure}

\section{Observation of Fermi arcs} In the underdoped cuprates and above the superconducting region, the state of matter is very unusual. In one hand, it is a metallic phase, confirmed by transport experiments. On the other hand, there is no closed Fermi surface. Along some directions and segments of the Fermi surface, excitations are gapless, while on other directions, excitations are gapped \cite{Kanigel_2006_a,Norman_2007a,Chubukov_2007a}. c-axis transport measurements also show a gap in the excitations that implies a bound state of Fermions. It has been discussed by Norman {\em et al} \cite{Norman_2007a} that if the scattering rate of quasiparticles $\gamma$ varies linearly with temperature, one can explain the observations. As we discussed in the previous paragraph, gauge fluctuations can cause such a behavior. Some authors \cite{Chubukov_2007a} have discussed that the electron phonon interaction may play an important role here and result in the linear $T$ dependence of $\gamma$ above some at high enough temperatures. Here we show that the interaction of slave particles with gauge bosons and the scattering of the quasiparticles from the d-wave potential results in such an exotic behavior. Using the method introduced in the previous section, we can compute the scattering rate of electrons in the superconducting region as well as the pseudogap phase. In the superconducting phase, the scattering rate is very small due to the condensation of both spinons and holons. In the pseudogap phase however, it is comparable to the pseudogap, varies linearly with $T$ and cannot be neglected. In this region we assume that there is a local pairing potential that electrons scatter off. This assumption leads to the following expression for the electrons Green's function:

\begin{eqnarray}
  G\para{k,\omega}^{-1}=\omega-\epsilon_{k}+i\gamma - \frac{\Delta_{k}^2}{\omega+\epsilon_{k}+i\gamma }
\end{eqnarray}
where, $\gamma\para{T}$ is the scattering rate at temperature $T$, energy of free quasiparticles. We have assumed d-wave pairing, {\em i.e.} $\Delta_{k} \para{T}=\Delta\para{T}\para{\cos\para{k_x}-\cos\para{k_y}}$. Using the above expression we can compute the spectral function and from the position of its peak we can read the energy of interacting quasiparticles. When $\epsilon_{k}=0$, it can be shown that as long as $\gamma >\sqrt{3}\Delta_{k}\para{T}$, the maximum is at $E=0$ and as a result we have gapless excitations along that direction. At nodal points $\Delta_{k}=0$ and we can always satisfy this equation at that point. Since $\gamma\para{T}\ll \Delta\para{T}$ in the superconducting phase, this condition is only satisfied very near nodal points and the length of Fermi arcs is negligible.  If we parameterize $\Delta_{k}$ by the angle, we have $\Delta\para{\phi,T}=\Delta\para{T}\cos\para{2\phi}$, and nodal points are located at $\phi=\left\{\pm \pi/4, \pm 3\pi/4\right\}$. Therefore the length of arc is $L\para{T}=8\sin^{-1}\para{\gamma\para{T}/{\sqrt{3}\Delta\para{T}}}$. The crossover temperature of the pseudogap phase, $T^*$, can be found by solving  $\gamma\para{T^*}=\sqrt{3}\Delta\para{T^*}$. Fig. 2 summarizes our results.

In conclusion, we reexamined the effect of gauge fluctuations on the single particle scattering rate. Using the mass of spinons proportional to the doping and doing a self-consistent calculation for the self-energy of quasiparticles scattering off the gauge field, we found the linear temperature dependent transport scattering rate of electrons above a crossover temperature that scales with the doping. Below this crossover temperature, we obtained $T^2$ behavior for the transport scattering rate that recovers the Fermi liquid behavior. We also found a scattering rate that is linear in $T$ above the crossover temperature. We showed that the linear dependence of the scattering rate explains the existence of the Fermi arcs in the pseudogap phase. The success of our model emphasizes on the importance of including gauge fluctuations in understanding the underlying physics of cuprates. We predict that the Fermi velocity in the Fermi liquid phase varies linearly with the doping.

{\bf Acknowledgements.---}  I thank X.G. Wen for financial support and very helpful discussions. I acknowledge useful discussions with P.A. Lee, T. Senthil, R. Asgari, J. McGreevy, B. Swingle, K. Michaili and R. Flint.

\bibliography{LTR}
\end{document}